# Non-reciprocal Optical Mirrors Based on Spatio-Temporal Modulation

R. Fleury[1,2], D. L. Sounas[1], and A. Alù[1,*]

[1]Department of Electrical & Computer Engineering, The University of Texas at Austin, Austin TX 78712, USA

[2]Laboratory of Wave Engineering, Institute of Electrical Engineering, EPFL, Station 11, 1015 Lausanne, Switzerland

*Correspondence to: alu@mail.utexas.edu

*The recent surge of interest in temporal modulation schemes to induce magnet-free non-reciprocity has inspired several exciting opportunities for photonic technology. Here, we investigate a scheme to realize free-space isolators and highly non-reciprocal mirrors with weak modulation imparted by an acoustic wave. Conventional optical mirrors are reciprocal: in a given plane of incidence, reflection is independent of the sign of the angle of incidence, which enables two people to simultaneously look at each other through their reflection. In contrast, we propose a strategy to dramatically break this symmetry by exploiting resonant interactions between a travelling acoustic wave and highly resonant guided optical modes, inducing total reflection of an optical beam at a given angle, and no reflection at the negative angle. Different from conventional acousto-optic isolators, which are based on non-resonant frequency conversion and filtering, our proposal operates at the frequency of the optical signal by tailoring the resonant properties of the structure as well as the acoustic wave frequency and intensity, enabling 50 dB isolation with*

*modest modulation requirements. Operation in reflection allows for close-to-zero insertion loss, enabling disruptive opportunities in our ability to control and manipulate photons.*

1. **Introduction**

One of the most fundamental properties of optical systems is reciprocity, according to which transmission of light between two points of space is identical for both transmission directions [1]-[5]. If light can propagate from a source to an observer following a given path in a complex environment, the reversed propagation path is also possible, and light transmission through forward and reversed paths is identical. Reciprocity plays an important role in the design and analysis of optical systems, including the symmetry between the transmission and reception properties of optical antennas, the bi-directionality of communication channels, and the symmetry between absorption and thermal emission of hot bodies. Nonetheless, there is a large number of situations in which this property is not desirable, and one would like to force light to travel along one-way paths. For instance, when a high-power laser source is fed to an optical set-up, reflections at interfaces with optical components can be routed back to the source, hamper its behavior and damage it. This problem can be prevented by connecting isolators at the output ports of the lasers, which allow transmission of signals from the laser to the optical setup but not in the opposite direction [6]. Other relevant nonreciprocal devices are circulators, which can separate signals propagating in opposite directions [7],[8], and can be used to realize topological insulators for light, which support unimpeded light transmission around sharp corners or other discontinuities along the signal path over continuous bandwidths [9].

Reciprocity in optics is described through the Lorentz reciprocity theorem, which holds for linear time-invariant (LTI) structures, consisting of materials with symmetric permittivity and permeability tensors ($\bar{\bar{\varepsilon}}^T = \bar{\bar{\varepsilon}}$ and $\bar{\bar{\mu}}^T = \bar{\bar{\mu}}$) [1],[2],[10]. The conventional approach to break reciprocity is to employ magneto-optical materials, such as iron garnets, breaking time-reversal symmetry under biasing with static magnetic fields [11]-[17]. However, this approach is accompanied by several drawbacks, including scarcity of magnetic materials, weak nonreciprocal response leading to large devices and incompatibility with integrated technologies. According to the Onsager principle of microscopic reversibility, LTI nonreciprocity can be obtained through biasing with any quantity that is odd-symmetric under time reversal [3],[4]. Apart from the magnetic field, the electrical current and the linear and angular momentum have this property. Current-based nonreciprocity can be obtained with transistors, which acquire nonreciprocal characteristics when biased with a static electric current [18]-[21]. However, the inherent nonlinearity and poor noise characteristics of transistors make these devices unsuitable for applications requiring handling of very strong or very weak signals, as in the majority of communication systems. Furthermore, this approach is limited to microwave and millimeter-wave frequencies, where transistors are available. Momentum biasing has been recently successfully applied in acoustics, by imparting a constant air flow to acoustic channels, such as hollow ring cavities [22]-[24]. This was a big step in acoustics, where nonreciprocal devices have been missing due to the very weak nonreciprocal response of acousto-magnetic materials. However, extension of the same concept to optics is challenging, due to the much larger propagation velocity of optical waves compared to sound, which results in practically unrealistic velocities for the moving media.

The previous limitations can be overcome by lifting the LTI assumption of the reciprocity theorem. Nonlinear effects combined with spatial asymmetries can lead to strong nonreciprocal responses, but with limitations in terms of the power of the incident signals and for simultaneous excitation of the structure from different ports [25]-[33]. On the other hand, temporally modulated structures can lead to strong linear nonreciprocal response and for this reason they have attracted a lot of interest during the past few years. Different approaches have been explored in this context. For example, one option is to use spatiotemporal modulation of a structure with the form of a traveling wave in order to effectively impart linear or angular momentum, which according to the Onsager principle breaks reciprocity [34]-[47]. Another possibility is to exploit the effective gauge field that appears in any frequency conversion process and design nonreciprocal devices based on arrays of optical mixers or modulated resonators [48]-[56]. Optomechanical effects can also lead to optical nonreciprocity, where the optical properties of a structure are modulated in space and time through a mechanical mode [57]-[70]. All these approaches have been used to demonstrate theoretically and experimentally a plethora of nonreciprocal devices, including isolators, circulators, nonreciprocal phase shifters and topological insulators, over different parts of the spectrum, ranging from microwaves to optics, and even extending to the quantum regime [71]-[74].

Here we introduce a scheme to realize a nonreciprocal mirror based on weak spatiotemporal modulation, which is able to reflect waves from a particular direction but not the complementary one. Our approach is based on imparting a traveling-wave spatiotemporal modulation to a Fabry-Pérot (FP) cavity through the excitation of an acoustic wave that breaks the symmetry of the cavity modes propagating in opposite directions, which are excited by incident waves from complementary directions. We develop an analytical model for the structure based on coupled-mode theory, and derive the conditions that lead to unitary and almost zero reflection from complementary directions. The proposed structure overcomes the insertion-loss challenges existing in nonreciprocal devices operating in transmission and it can be useful for the design of efficient free-space optical isolators.

## 2. Nonreciprocal optical mirror

Consider an optical signal emitted from the left and being received on the right after bouncing-off a mirror (Figure 1a, green ray). Since Maxwell's equations are invariant under the time-reversal operation, the oppositely propagating wave (purple ray), with reversed propagation path, is also a viable electromagnetic solution. If we are able to slightly change the properties of the mirror, such that its time-reversed image is no longer identical, we may design the structure such that the purple ray is still reflected with the same large reflectance, whereas the green ray goes through. In such a configuration (Figure 1b), one of the observers can see the other without being seen, therefore breaking reciprocity, and realizing an isolator in reflection with low insertion loss.

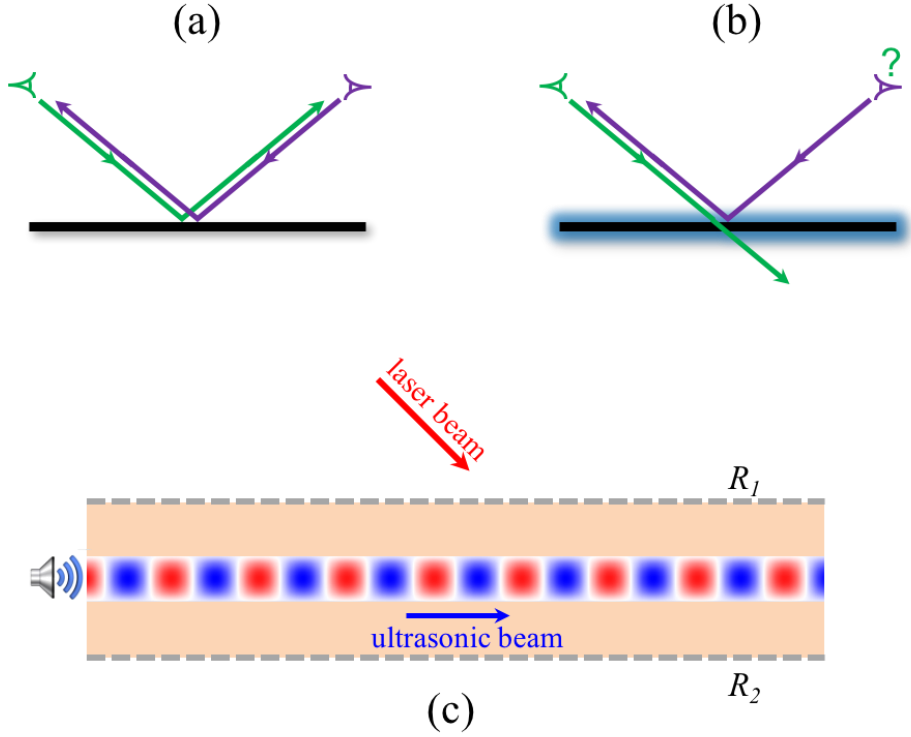

**Figure 1: The concept of non-reciprocal mirrors.** (a) With a reciprocal mirror, two observers can simultaneously look at each other through the mirror. (b) A nonreciprocal mirror allows one observer to observe the second, but not the opposite. (c) Proposed scheme to achieve a non-reciprocal mirror: A laser beam is incident on a Fabry-Pérot resonator built from an acousto-optical material sandwiched between two partially reflective surfaces (in grey, with reflectance $R_1$ and $R_2$).

To achieve this much-sought functionality, we consider a FP cavity composed of a thick dielectric slab of acousto-optic material, sandwiched between two partially reflecting surfaces of reflectance $R_1$ and $R_2$, whose values are close to unity. The cavity supports forward and backward propagating modes $|+k_t\rangle$ and $|-k_t\rangle$ with opposite transverse wavenumbers $k_t$ and $-k_t$, respectively. These modes have the same frequency $\omega_0$, as a result of reciprocity. A monochromatic optical plane wave is incident on the structure at a fixed angle $\theta_i$ which matches the transverse wavenumber of the cavity modes, in particular $k_t = k_0 \cos\theta_i$, where $k_0$ is the wavenumber in free space. In addition, we send an ultrasonic wave along the slab, effectively imparting a travelling refractive index modulation in the resonator $n(y,t) = n + \Delta n \cos(\omega_m t - k_m y)$, where $k_m$ is the modulation wavevector and $\Delta n/n \ll 1$. By operating at the Brillouin scattering condition $\boldsymbol{k}_m = 2\boldsymbol{k}_t$, this modulation induces photonic transitions between the forward $|+\boldsymbol{k}_t\rangle$ and backward $|-\boldsymbol{k}_t\rangle$ cavity modes. To formally describe this effect, we can expand the photonic state excited in the cavity $|\psi(t)\rangle$ in the basis of forward and backward modes of the unperturbed cavity as $|\psi(t)\rangle = a_+(t)|+\boldsymbol{k}_t\rangle + a_-(t)|-\boldsymbol{k}_t\rangle$. Using perturbation theory [75], we obtain the set of coupled-mode equations

$$\dot{a}_\pm = \left(-i\omega_0 - \tau_1^{-1} - \tau_2^{-1}\right)a_\pm + i\omega_0 \frac{\Delta n}{2n} e^{\mp i\omega_m t} a_\mp + \sqrt{2\tau_1^{-1}} s_\pm^{in}. \tag{1}$$

In Eq. (1), $\tau_1^{-1}(\tau_2^{-1})$ are the decay rates at the interface of reflectance $R_1$ ($R_2$) for the cavity modes, which is the same for both propagation directions due to reciprocity. The quantity $s_+^{in}$ ($s_-^{in}$) corresponds to the complex amplitude of the incident field along (against) the direction of propagation of the $|+\boldsymbol{k}_t\rangle$ ($|-\boldsymbol{k}_t\rangle$) mode. Following coupled-mode theory [76], the outgoing signals are $s_\pm^{out} = -s_\pm^{in} + \sqrt{2\tau_1^{-1}} a_\pm$. We can solve (1) assuming monochromatic excitation with unitary amplitude $s_\pm^{in} = e^{i\omega t}$, and calculate the reflection coefficients from left to right $r_+$, and from right to left $r_-$, corresponding to excitation along or against the modulation momentum. Defining $K_m = \Delta n/(2n)$ as the modulation depth, $Q = \omega_0/2(\tau_1^{-1} + \tau_2^{-1})$ as the FP quality factor, and $\Omega_i = \omega_i/\omega_0$ as the normalized frequencies, we obtain

$$r_\pm = \frac{(1-\Omega+iQ^{-1}T)(\Omega-1\mp\Omega_m+iQ^{-1}/2)+K_m^2}{(\Omega-1+iQ^{-1}/2)(\Omega-1\mp\Omega_m+iQ^{-1}/2)-K_m^2}. \tag{2}$$

In addition, $T = (q-1)/(2q+2)$, where $q = T_1/T_2$ is the coefficient of asymmetry of the cavity, with $T_{1,2} = 1 - R_{1,2}$. From Eq. (2), we see that in absence of acoustic modulation, ($K_m = 0$), we necessarily get symmetric reflection ($r_+ = r_-$), a consequence of time-reversal symmetry. However, as soon as we turn on a time-dependent modulation ($K_m \neq 0$ and $\Omega_m \neq 0$), breaking time-reversal symmetry, we induce the desired non-reciprocity in reflection, $r_+ \neq r_-$. The incidence angle for which nonreciprocity is maximum is derived from $\boldsymbol{k}_m = 2\boldsymbol{k}_t$ as

$$\boldsymbol{\theta}_i = \cos^{-1}\left(\frac{c}{2v}\frac{\omega_m}{\omega_i}\right), \quad (3)$$

where $c$ is the speed of light in free space and $v$ is the velocity of the acoustic wave.

While it is not necessarily surprising that we can break time-reversal symmetry imparting a synthetic linear momentum to the resonator, a question of crucial importance is whether we can practically induce large non-reciprocity in reflection with this setup, considering a moderate modulation frequency (hundreds of MHz), orders of magnitude below the optical signal frequency (hundreds of THz), and with a small modulation depth typically achievable via acousto-optical interactions ($\Delta n/n = 10^{-6}$ to $10^{-5}$) [77],[78]. To answer this question, we study the mathematical conditions under which the reflection coefficients $r_\pm$ can reach extremely small values, exploiting Eq. (2). We find that reflectance minima occur at specific frequencies, always close to $\omega_0$, however these frequencies are different for $r_+$ and $r_-$ and separated by $\Delta\omega = \sqrt{\omega_m^2 + 4\omega_0^2 K_m^2}$. These minima occur at $\omega_\alpha = \omega_0 - \Delta\omega/2$ for $r_+$, and at $\omega_\beta = \omega_0 + \Delta\omega/2$ for $r_-$. Therefore, $\omega_\alpha$ and $\omega_\beta$ are always the frequencies for which non-reciprocal response is the strongest. By evaluating $r_\pm$ at $\omega_\alpha$, we find that $r_+$ and $r_-$ only depend on two dimensionless parameters, $QK_m = Q\delta n/2n$ and $Q\omega_m/\omega_0$. For a 8 mm thick TeO$_2$ slab, with $n = 2.26$, $R_1 = 0.96$ and $R_2 = 0.98$, standard formulas [79] give $Q = 2.5 \times 10^6$, and we can practically reach values of $QK_m$ up to 12 [77]. Notice that this value of $Q$, which appears large for nanophotonic structures, is easily achievable in a mm-thick Fabry-Perot resonator, given the large electrical volume. For operation at 200 THz (optical wavelength of 1.5 μm), $Q\,\omega_m/\omega_0$ can therefore reach values up to 10 without particular technological challenges. Accordingly, in Figure 2 we fix $QK_m = 5$ and plot the magnitude of $r_+$ and $r_-$ at the frequency $\omega_\alpha$, as a function of the parameter $Q\,\omega_m/\omega_0$. Panel (a) is obtained assuming a symmetric cavity, i.e., for $q = 1$. We see that non-negligible non-reciprocal reflection occurs for values of $Q\,\omega_m/\omega_0$ above unity. At $Q\,\omega_m/\omega_0 = 10$, we get 20 dB isolation in reflection, with unitary backward reflection, i.e., zero insertion losses. The case of an asymmetric cavity ($q = 2$, corresponding, for example, to $R_1 = 0.96$ and $R_2 = 0.98$) is shown in panel (b). Remarkably, for $Q\,\omega_m/\omega_0 = 6$, we reach a sweet spot for which $r_-$ is unitary whereas $r_+$ is smaller by 40 dB. This demonstrates that very good isolation can be reached with asymmetric non-reciprocal mirrors, with the clear advantage of zero insertion loss while maintaining large isolation levels.

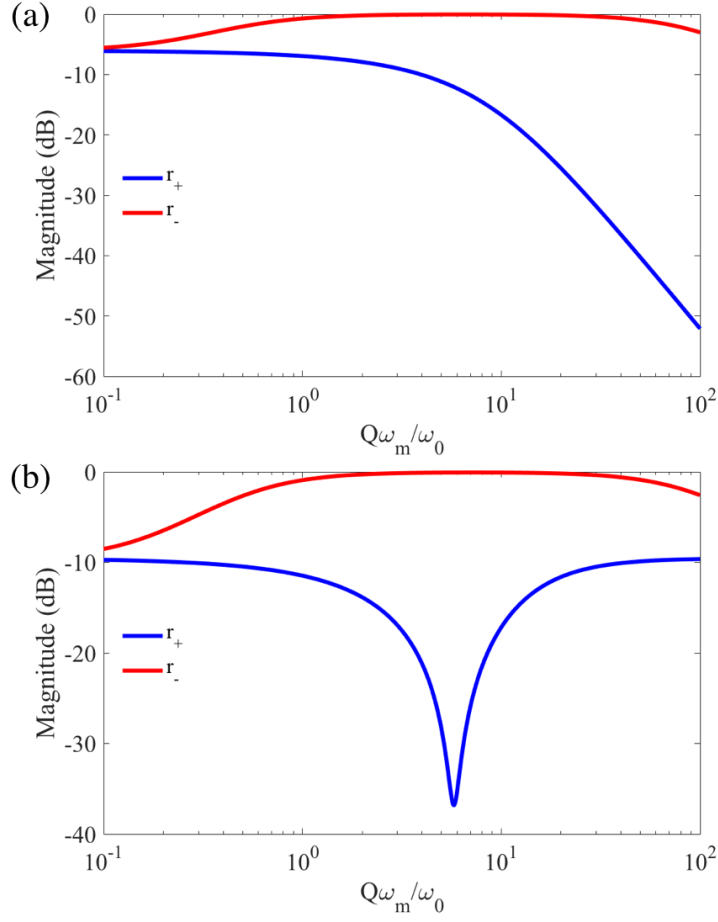

**Figure 2: Demonstration of non-reciprocal reflection** at the design frequency, for $Q\delta n/2n = 5$. Magnitude of the reflection coefficients for incidence along $(r_+)$ and against $(r_-)$ the direction of acoustic modulation, in decibels, as a function of the parameter $Q\,\omega_m/\omega_0$. Panel (a) considers a symmetric cavity ($q = 1$), whereas panel (b) corresponds to a slightly asymmetric cavity, with $q = 2$.

In order to get more physical insight into this exotic asymmetric reflection phenomenon, we plot in Figure 3a the frequency dependence of the reflection, targeting the optimal point of Figure 2b. The acoustic wave is at 456.3 MHz, and it induces a modulation of the TeO$_2$ refractive index $\delta n/n = 4 \times 10^{-6}$. With a speed of sound in TeO$_2$ of 4260 m/s, the Brillouin condition can be achieved at relatively low incidence angles, close to 5 degrees, as it can be found from Eq. (3).

The scattering spectrum features deep reflection dips that occur at frequencies very close to $\omega_\alpha$ and $\omega_\beta$, as expected. Remarkably, the blue curve (corresponding to $r_+$), differs by more than 50 dB from the red curve (corresponding to $r_-$) at its minimum. The isolation level is extremely high, while at the same time insertion loss is zero, confirming the ideal behavior of the non-reciprocal mirror.

To better grasp the physics underlying the scattering signature seen in Figure 3a, we calculate the optical eigenstates of the FP cavity. Using Eq. 1, we find that the slab supports the following source-free solutions, written in the $|\pm k_t\rangle$ basis ($C_1$ and $C_2$ are constants):

$$|\psi(t)\rangle = C_1\left\{e^{-i\omega_\alpha t}|+k_t\rangle + \frac{\Delta\omega}{2\omega_0 K_m}e^{-i(\omega_\alpha-\omega_m)t}|-k_t\rangle\right\} + C_2\left\{e^{-i\omega_\beta t}|-k_t\rangle - \frac{\Delta\omega}{2\omega_0 K_m}e^{-i(\omega_\beta+\omega_m)t}|+k_t\rangle\right\}. \quad (4)$$

In absence of modulation, Eq. (4) collapses to a state composed of two degenerate eigenmodes $|\pm k_t\rangle$ oscillating at $\omega_0$, as expected. The eigen-states of the time-dependent system are themselves composed of a forward $|+k_t\rangle$ and a backward $|-k_t\rangle$ component, oscillating at different frequencies (Figure 3b). When light is incident along the modulation (case of $r_+$, blue curve), it can only couple to the $|+k_t\rangle$ component of the eigenstates of the system. This happens at $\omega_\alpha$ and $\omega_\beta + \omega_m$, consistent with the dips in Figure 3a. Similar considerations hold for the red curve.

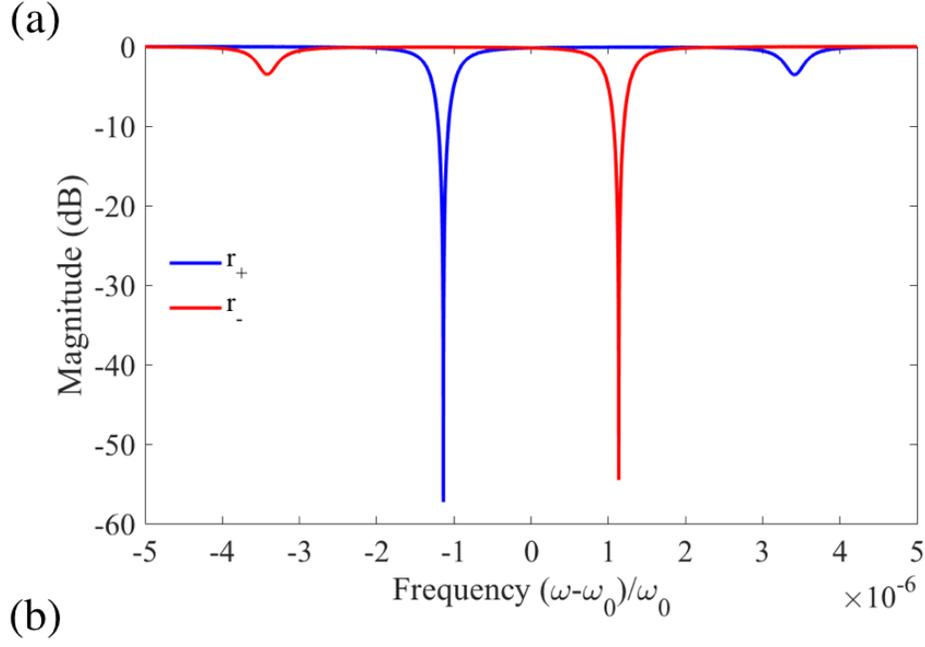

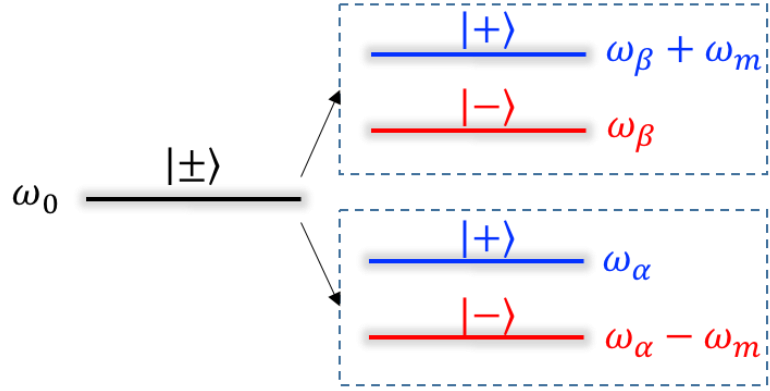

**Figure 3: Spectrum of the modulated FP cavity** for signals incident along ($r_+$) and against ($r_-$) the acoustic wave direction of propagation. The dips in the reflection spectrum (a) correspond to resonant tunneling condition at the specific eigenfrequencies of the time-driven slab (b). We assume an optical plane wave (wavelength of 1.5 **μm**, i.e. $f_0 = \omega_0/2\pi = 200$ THz), incident on a 8 mm thick TeO$_2$ slab with $R_1 = 0.96$ and $R_2 = 0.98$. The acoustic signal is at 456.3 MHz, creating a relative optical refractive index variation $\delta n/n = 4 \times 10^{-6}$.

The effect of varying the angle of incidence, i.e., exciting the structure away from the Brillouin condition, is studied in Figure 4a. We plot the isolation, i.e., the ratio $|r_-|/|r_+|$ expressed in decibels, as a function of frequency and incidence angle. We see that the resonant tunneling conditions occur at different frequencies depending on the angle of incidence, and high isolation regions correspond to narrow angular windows. In panel (b), we quantify the asymmetric reflection obtained in the case of a 2D Gaussian beam, incident at the design angle, as a function of its diameter (defined as the full width at half maximum). We define the beam reflectance for incidence along (against) the modulation, denoted as $P_+$ ($P_-$), as the ratio between the total power carried by the reflected beam to the one carried by the incident beam. As expected, for sufficiently large beam diameters, the reflection is close to 100% for incidence against the modulation direction, and 0% at the specular angle. This demonstrates the unique capability of the proposed non-reciprocal mirrors to induce large asymmetry in reflection with close to zero insertion losses, even when the incident field is not a pure plane wave.

## 3. Conclusions

In summary, we have studied the possibility to build drastically non-reciprocal mirrors, that are capable of 100% reflection at a positive angle and 0% reflection at the negative angle, based on a weak form of spatio-temporal modulation imparted by an acousto-optical wave traveling along a resonant Fabry-Perot resonant cavity. Our results demonstrate the realistic possibility of achieving full optical isolation in reflection in a practical setup. Remarkably, operation in reflection allows for zero insertion loss. We stress that the present method is based on totally different physics than traditional acousto-optical modulators [23],[24], which are operated in transmission and are inherently non-resonant, surrounded by anti-reflection coatings. These systems rely on completely different phenomena, exploiting frequency conversion and filtering.

While conventional mirrors are intrinsically constrained by time-reversal symmetry, we envision that the proposed non-reciprocal mirrors may enable the realization of disruptive photonic technology and open a wealth of new opportunities in our ability to control and guide visible light.

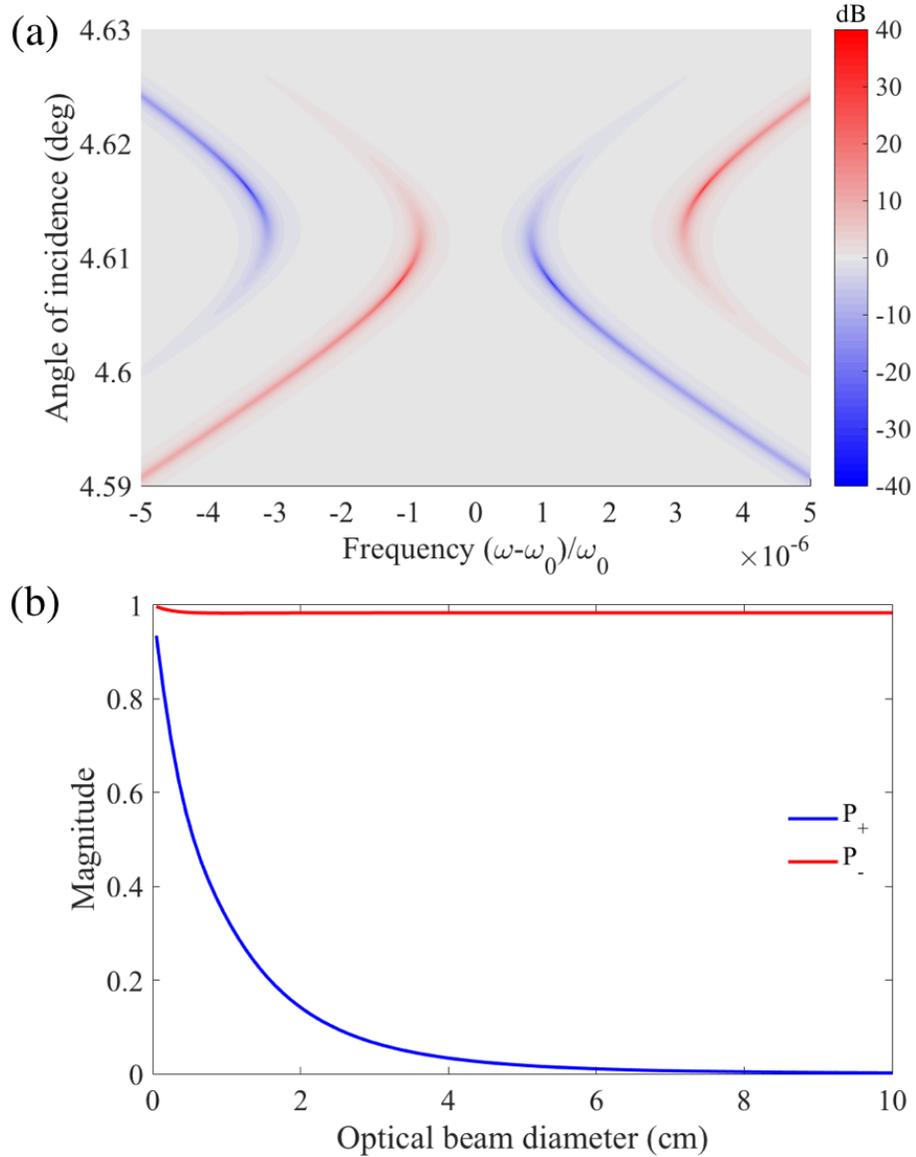

**Figure 4: Effect of the angle of incidence and optical beam size.** (a) Isolation of the mirror, defined as $|r_-|/|r_+|$, as a function of the frequency of the optical signal and its angle of incidence. (b) Total transmitted power for a 2D Gaussian beam as a function of its beam diameter at 200 THz. The beam is centered around an incidence angle of 4.61 degrees.

**Acknowledgments**

This work was supported by the National Science Foundation. The authors are grateful to Prof. M. Belkin for useful suggestion.